\documentclass[12pt]{article}
\usepackage{latexsym}
\usepackage{epsfig}
\newcommand{\vm}{V_{m\bar{m}}(R)}
\newcommand{\beq}{\begin{equation}}
\newcommand{\eeq}{\end{equation}}
\newcommand{\beqn}{\begin{eqnarray}}
\newcommand{\eeqn}{\end{eqnarray}}
\newcommand{\bea}[1]{\beq\begin{array}{#1}}
\newcommand{\eea}{\end{array}\eeq}
\newcommand{\eq}[1]{(\ref{#1})}

\newcommand{\dual}[1]{{}^{*}{#1}}

\newcommand{\dd}{{\mathrm d}}
\newcommand{\Z}{Z\!\!\! Z}

\newcommand{\diff}{\partial}

\newcommand{\cZ}{{\cal Z}}

\newcommand{\dD}{{\cal D}}

\newcommand{\NP}[3]{{\it Nucl. Phys. }{\bf #1} (#2) #3}
\newcommand{\NPPS}[3]{{\it Nucl. Phys. Proc. Suppl. }{\bf #1} (#2) #3}
\newcommand{\PL}[3]{{\it Phys. Lett. }{\bf #1} (#2) #3}
\newcommand{\PRL}[3]{{\it Phys. Rev. Lett. }{\bf #1} (#2) #3}
\newcommand{\PRep}[3]{{\it Phys. Rep. }{\bf #1} (#2) #3}
\newcommand{\PR}[3]{{\it Phys. Rev. }{\bf #1} (#2) #3}

\newcommand{\IJMP}[3]{{\it Int. J. Mod. Phys. }{\bf #1} (#2) #3}

\begin{document}

\date{}
\title{On the heavy monopole potential in gluodynamics.
\vskip-40mm
\rightline{\small MPI-PHT-00-50}
\vskip 40mm }

\author{M.N. Chernodub$^{\rm a}$, F.V.~Gubarev$^{\rm a,b}$,
        M.I. Polikarpov$^{\rm a}$,V.I.~Zakharov$^{\rm b}$ \\
\\
$^{\rm a}$ {\small\it Institute of Theoretical and  
	Experimental Physics,} \\
           {\small\it B.Cheremushkinskaya 25, Moscow, 117259, Russia}\\
$^{\rm b}$ {\small\it Max-Planck Institut f\"ur Physik,}\\ 
           {\small\it F\"ohringer Ring 6, 80805 M\"unchen, Germany}\\
}
\maketitle
\begin{abstract}
\noindent
We discuss predictions for the interaction energy of the fundamental monopoles
in gluodynamics introduced via the 't~Hooft loop. At short distances, the heavy
monopole potential is calculable from first principles. At larger distances, we
apply the Abelian dominance models. We discuss the measurements which would
be crucial to distinguish between various models. Non-zero temperatures are also
considered. Our predictions are in qualitative agreement with the existing lattice data.
\end{abstract}

%====================================================================
\section{Definition of the heavy monopole potential}

In this note we discuss the interaction of the          
fundamental monopoles in gluodynamics.
The fundamental monopoles can be introduced via the 't~Hooft loop \cite{'tHooft:1978hy},
and are best understood on the lattice. The monopoles correspond
to point-like objects in the continuum limit and can be visualized as end-points of the
Dirac strings which in turn are defined as piercing negative plaquettes.
In more detail, consider the standard Wilson action of $SU(2)$ lattice gauge theory (LGT):
\beq
S_{lat}(U) = - \beta\sum\limits_p \; \frac{1}{2}\mathrm{Tr}\;
U_p\,,
\label{plaquette}
\eeq
where $\beta=4/g^2$, $g$ is the bare coupling, the sum is taken over all
elementary plaquettes $p$ and $U_p$ is the ordered product of link matrices
in the fundamental representation along the boundary of $p$. Then the 't~Hooft loop
is formulated (see, e.g. \cite{Hoelbling:2000su,forcrand} and references therein) in terms of
a modified action $S(\beta,- \beta)$:
\beq
S(\beta,- \beta) =
- \beta \sum\limits_{p\notin \dual \Sigma_j} \;
\frac{1}{2}\mathrm{ Tr}\;U_p
+ \beta\sum\limits_{p\in \dual \Sigma_j}\;\frac{1}{2}\mathrm{ Tr}\;U_p\,,
\label{S1}
\eeq
where $\dual \Sigma_j$ is a manifold which is dual to a surface spanned on the monopole
world-line $j$. Introducing the corresponding partition function,
$Z(\beta,-\beta)$ and considering a time-like planar rectangular $T\times R$, $T\gg R$
contour $j$ one can define
\beq
V_{m\bar{m}}(R) ~\equiv ~ -\frac{1}{T} \ln{ Z(\beta,-\beta)\over Z(\beta,\beta)}\,.
\label{energy}
\eeq

Since the external monopoles become point-like
particles in the continuum limit the potential $V_{m\bar{m}}(R)$ is
the same fundamental quantity as, say, the heavy-quark potential $V_{Q\bar{Q}}$
related to the expectation value of the Wilson loop.
By analogy,
we will call the quantity $V_{m\bar{m}}(R)$ the heavy monopole potential.
 Because of the asymptotic freedom, one would expect that at short distances the potential
$V_{m\bar{m}}(R)$ is predictable from first principles. Some preliminary work is needed,
however, since originally $V_{m\bar{m}}(R)$ is formulated in the lattice terms.
The continuum version of the 't~Hooft loop was worked out in \cite{gubarev1}.
Moreover, it turns possible \cite{gubarev2} to reformulate the problem in the Lagrangian
approach a la Zwanziger \cite{zwanziger}. The crucial point is that the role of the Dirac
string, explicit in the lattice construction (\ref{S1}), reduces to particular rules of
the ultraviolet regularization. The papers \cite{gubarev1,gubarev2} set up a theoretical framework
which we will exploit here.

Recently, first direct  measurements of $V_{m\bar{m}}(R)$ on the lattice
were reported \cite{Hoelbling:2000su,forcrand}. The main conclusion is that at large distances
the potential $V_{m\bar{m}}(R)$ is of the Yukawa type.
Motivated by these measurements we explore in this note how far one can reach
with predicting the heavy monopole potential theoretically. As is already mentioned above,
we are not trying to develop new models but rather to elucidate the relation between various
theoretical hypotheses  and measurable properties of the potential $V_{m\bar{m}}(R)$.

%====================================================================
\section{Potential at zero temperature}

\subsection{Coulomb-like potential}

Consider first the potential $\vm$ at short distances. Then the interaction between the
monopoles is Coulomb-like (see \cite{gubarev1,gubarev2} and references therein):
\beq
\label{small-R-potential}
V_{m\bar{m}}(R) ~=~ - {\Biggl(\frac{2 \pi}{g(R)}\Biggr)}^2 \, \frac{1}{4 \, \pi \, R}
~ = ~ - {\pi\over g^2(R)} \;{1\over R}\,,
\eeq
where $g(R)$ is the running coupling of the gluodynamics. Eq.~(\ref{small-R-potential}) makes
manifest that monopoles in gluodynamics unify Abelian and non-Abelian features. Namely, the overall
coefficient, $\pi/g^2$ is the same
as in the Abelian theory with fundamental electric charge $e=g$ while the running
of the coupling $g^2$ is governed by the non-Abelian interactions.

Prediction (\ref{small-R-potential}) is rooted in the very foundations
of the theory. Namely, the $U(1)$ overall normalization is based on the classification of the monopoles
in non-Abelian gauge theories (see, e.g., \cite{coleman} and references therein). The running
of the coupling reflects the general rule that the effect of the fluctuations at short distances
can be absorbed into the renormalization of the coupling. What might be not so evident, is
which distances can be considered small in practice. We shall come to discuss this point later.

To derive (\ref{small-R-potential}) explicitly one establishes first the classical equations of motion
corresponding to the 't~Hooft loop:
\beq
\label{landau}
 \diff^2 A_{\mu}^a~=~ - \frac{2\pi}{g}\partial_{\nu}\, \big( n^a \, \dual{\Sigma}_{\mu\nu}\big)\,,
\eeq
where $\Sigma_{\mu\nu}$ is the world sheet of the Dirac string whose end points represent the monopoles,
star symbol means the duality operation and $n^a$ is a unit vector in the color space\footnote{
%-----------------------------------------------
For simplicity we consider only the $SU(2)$ gauge group.
%-----------------------------------------------
} whose direction is a matter of gauge fixing. In particular, there exist gauges in which
$n^a$ is constant.

Eq.~(\ref{landau}) demonstrates that only the boundary of the Dirac string determines the
interaction. Therefore, one can hope to introduce Lagrangian which would reproduce the same
results in terms of particles alone, without strings. A well known example of such a construction is
the Zwanziger Lagrangian \cite{zwanziger} which allows to describe photons interacting with
both magnetic and electric charges. In case of gluodynamics 
the Zwanziger-type Lagrangian was found in \cite{gubarev2}:
\beq
\label{b-action}
L_{\mathrm{ dual}}(A^a,B) ~=~ \frac{1}{4}(F^a_{\mu\nu})^2
~+~ \frac{1}{2}[m\cdot(\partial\wedge B - i \dual{n^a F^a})]^2 ~+~ i \frac{2\pi}{g} j B\,,
\eeq
where $j$ is the magnetic current, $F_{\mu\nu}^a$ is the non-Abelian field strength tensor and
$m_\mu$ is an arbitrary constant vector, $m^2=1$.
Note that the vector $m$ enters also the original Zwanziger Lagrangian
in the QED case. The vector field $B$ can be called dual gluon,
although because the $A^a$ and $B$ fields are mixed
the number of the physical degrees of freedom corresponding to (\ref{b-action})
is the same as in the gluodynamics itself.
One can demonstrate that the Lagrangian (\ref{b-action}) reproduces
the heavy monopole potential (\ref{small-R-potential}).
The Dirac string, however, re-emerges in a way, through rules of the ultraviolet
regularization which uniquely define the theory (\ref{b-action}) on
the quantum level.

A salient feature of the Eq.~(\ref{b-action}) is that the dual gluon $B$
is a $U(1)$ particle despite of the fact that the ``ordinary'' gluon are
in an adjoint representation of the non-Abelian group. There is no violation, however, of
the color symmetry through the mixing of the $B$ and $A^a$ fields since the choice of $n^a$
is a matter of gauge fixing, see for details \cite{gubarev1,gubarev2}. In the simplest
form, the argument is that an averaging over all possible embeddings of the $U(1)$ into
the non-Abelian group is understood. Thus, by checking (\ref{small-R-potential})
one would verify that the dual gluon in gluodynamics is indeed a $U(1)$ gauge boson.

The message so far is that at one-loop level the interaction of the heavy monopoles is
understood theoretically no worse than the interaction of the heavy quarks.
There is a word of caution, however, concerning higher loops. The point is that
the monopole field is proportional to $g^{-1}$. As a result, any number
of interaction of virtual gluons with the external monopole field is equally
important and the evaluation of the corresponding determinant looks a
very difficult problem, with no analytical solution at all \cite{goebel}.
The leading log approximation is an exception to the rule \cite{gubarev2} and is
given by the simple graph with
insertion of two external fields. Thus, it may take long before the
next term in the expansion in $g^2(R)$ is found.

%====================================================================
\subsection{Power corrections}

As far as the physics of short distances is concerned, the next step is to consider power
corrections to (\ref{small-R-potential}). Theoretically, the prediction is that there is no
linear in $R$ term at small~$R$:
\beq
\label{small-R}
V_{m\bar{m}}(R) ~\approx~ - {\pi\over g^2(R)} \;{1\over R}~+~a_0\Lambda_{QCD}~+~a_2\Lambda_{QCD}^3R^2+...
\eeq
where $a_{0,2}$ are constants sensitive to the physics in the infrared.
To the contrary, the coefficients in front of the odd powers of $R$ are sensitive to
the short distances. The same  is true in case of the heavy quark potential
and we refer the reader to \cite{az} for the reasoning and further references.
The absence of the linear term follows then from the absence of a dimension
$d=2$ quantity in the gluodynamics.

Note that in case of the heavy quark potential this logic can in fact be challenged
\cite{ss}. Indeed, the power corrections are obviously sensitive to the vacuum structure.
The picture that we have in mind is that monopoles condense in the vacuum (see below).
Then the external quarks are connected by a Dirac string and since the string is
infinitely thin, the physics in the ultraviolet can change \cite{ss}.
The case that we are considering now, that is, magnetically charged external probes
embedded into the vacuum with monopoles condensed, is much simpler. Indeed, in the dual
language we do not need then Dirac strings at all. Thus, there is no mechanism to change
(\ref{small-R}). Things would change if we included light quarks.
Then the quark condensate could invalidate (\ref{small-R}) since the quarks are
carrying color. However, the dynamical quarks would also make visible the Dirac string
connecting the external monopoles, and the whole construction would radically change.
We would not consider that case in detail.

To summarize, the absence of the linear in $R$ term at short distances follows
also from the first principles and eventually is related to the fact that
the monopoles are not confined.
%====================================================================
\subsection{Yukawa potential}

It is a common point that at large distances $\vm$  is of the Yukawa type,
see, e.g., \cite{Hoelbling:2000su,forcrand,gubarev2,samuel,stack}:
\beq
\label{yukawa}
V_{m\bar{m}}(R) ~=~ - C \cdot {e^{-
\mu R}\over R}~.
\eeq
In fact, it is a manifestation of the general statement that either electric or magnetic
charges are confined \cite{thooft2}.
At a closer look, however, there is  a variety of model dependent predictions for
the parameters $C$ and $\mu$. Thus, experimental determination of these parameters could distinguish
between various models and we will discuss the models one by one:

{\it (i)} Continuing with our discussion of the {\it fundamental} gluodynamics,
Eq.~(\ref{small-R}) tells us that the Yukawa potential (\ref{yukawa}) cannot be true at all the distances.
Indeed, absence of the linear term is inconsistent with (\ref{yukawa}).
In other words, there is no reason to believe that the Yukawa potential matches smoothly
the Coulomb-like potential and, generally speaking, $C\neq \pi/g^2$. Thus, there are no model
independent predictions either for $\mu$ or $C$.

{\it ({ii})} Historically, the first prediction for $\mu$ was obtained in \cite{samuel}.
 Namely, it was shown that to any order in the {\it strong coupling expansion} the mass $\mu$
coincides with the mass of $0^{++}$ glueball:
\beq
\label{glueball}
\mu~=~m_G.
\eeq
As for the constant $C$, there is again no reason for it to be the same for the Yukawa and
Coulomb like potentials  so that, generally speaking, $C\neq\pi/g^2$.

{\it ({iii})} The Yukawa potential (\ref{yukawa}) at all the distances arises naturally
within effective field theories with monopole condensation\footnote{
%--------------------------------------------------
Note that the monopoles which condense are of course {\it not} the fundamental monopoles which
are introduced via the 't~Hooft loop as external probes. Instead, the monopoles ``living'' in
the QCD vacuum  have a double magnetic charge.
%--------------------------------------------------
}. To describe the condensation in the field theoretical language, one introduces
a new (effective) field $\phi$ which is interacting minimally with the dual gluon
(for review and references see, e.g., \cite{gubarev2,baker}).
Consider first the Lagrangian proposed in \cite{gubarev2}.
\beq
L_{eff}~=~L_{\mathrm{ dual}}(A^a,B)+S_{\mathrm{Higgs}}(B,\phi_m)\,.
\label{higgs}
\eeq
where $L_{\mathrm{ dual}}$ is
given by Eq.~(\ref{b-action}) and $L_{\mathrm{Higgs}}$ is the standard Higgs part of the
Abelian Higgs model action.
The vacuum expectation value of the Higgs, or monopole field is, of course, of order
$\Lambda_{QCD}$. Within this model,
\beq
\label{predictionhiggs}
\mu~=~m_V\,,\qquad C~=~{\pi\over g^2}\,,
\eeq
where $m_V$ is the mass of the vector field $B$ acquired through the Higgs mechanism.
An attractive feature of the model~\eq{higgs} is that it can incorporate the Casimir
scaling~\cite{gubarev2}.

It is worth emphasizing that the predictions (\ref{predictionhiggs}) differ substantially
from the cases ({\it i}), ({\it ii}) discussed above. First, $m_V$ may not coincide
with any glueball mass. This is not in contradiction with the general principles since there
is no spectral representation for the correlator of two magnetic currents $j$
(which are sources of the field $B$) \cite{gubarev1}. Indeed, although the currents do not
carry any color index explicitly they are not color singlets either
since their color orientation is a matter of a gauge fixing. Moreover
the monopole trajectory $j$ is to be understood
as the boundary of the Dirac-string world sheet.
The Dirac string is infinitely heavy in the continuum limit
so that all the intermediate states are in fact infinitely heavy.

As for the magnitude of $m_V$, the estimates usually give $m_V\approx 1\mathrm{~GeV}$,
see, e.g., \cite{ilgenfritz}. More generally, if it turns out that $\mu$ is indeed smaller
than the lowest glueball mass, it would be a serious argument in favor of the Abelian
dominance models. If the value of $\mu$ does not distinguish between the models, then it
would be crucial to check that the potential is vector like, as is predicted by
the model~(\ref{higgs}).

At very short distances the potential (\ref{yukawa}) should yield to the potential
(\ref{small-R}) obtained within the fundamental gluodynamics. However, there exist
various pieces of evidence that the effective theory (\ref{higgs}) is valid down
to such small scales that in practice it covers all the distances
available for the lattice measurements nowadays \cite{ss, shuryak}.
Clearly, a careful study of the $\vm$ could be crucial to confirm or reject these
speculations.

({\it iv}) It is worth emphasizing that the value $C=\pi/g^2$ is specific for the
model (\ref{higgs}) and is not true in a generic {\it Abelian dominance model}.
To use an analogy, consider the world with spontaneously broken chiral symmetry
of strong interactions and with weakly interacting vector bosons added. Then the model
(\ref{higgs}) would be analogous to assuming that the massless pions are coupled
to the $W$ bosons but do not interact directly with the nucleons. Thus,
we could have added to (\ref{higgs}) interaction of the Higgs bosons with the fundamental
monopoles. In other words, since the monopoles are condensed the monopole charge
of particles is not well defined and the constant $C$ is not constrained, generally speaking.

%====================================================================
\section{Non-vanishing temperatures}

\subsection{Dimensional reduction approach}

The screening mechanism at high temperatures is the Debye screening. As is
noted above the classical limit of the state created by 't~Hooft loop is
an Abelian monopole pair. Thus it is natural to evaluate the Debye mass
within the Abelian dominance models. Note that even if one assumes the Abelian
dominance, it is an open question which $U(1)$ is to be used to classify
the monopoles. At short distances, there is no such problem since the
coupling $g^2$ is small and the only subtle point is the averaging over
all the embeddings of the $U(1)$ into $SU(2)$. At large distances when
$g^2\sim 1$, different definitions of the $U(1)$ result in different
physics. We assume that the maximal Abelian gauge is used. 

Our estimates of the mass $\mu(T)$ \cite{gubarev1} at high $T$ are based on the
observation that the
Abelian model which corresponds to the high temperature gluodynamics is the 3D compact $U(1)$
theory.
Therefore, at high temperatures the screening mass $\mu$ coincides with the corresponding
Debye mass:
\beq
\label{m_D}
\mu^2 ~=~ m^2_D = 16\pi^2 {\rho \over e^2_3},
\eeq
where $\rho$ is the density of monopoles and $e_3$ is the corresponding
three-dimensional coupling constant.
To estimate the temperature dependence of $m_D$ we use
the numerical results of Ref.~\cite{Bornyakov}, where the density of Abelian monopoles was
obtained\footnote{
Note that the original result of Ref.~\cite{Bornyakov} 
refers to the {\it lattice} monopole density
$\rho_{lat.} = 0.50(1) \, \beta^3_G$, where $\beta^3_G$ is a three dimensional coupling
constant which is expressed in terms of the $3D$ electric charge $e_3$ and lattice spacing $a$
as  $\beta^3_G = 4 \slash (a \, e^2_3)$. The {\it physical} density $\rho$ of monopoles is given by
$\rho=\rho_{lat.} \, a^{-3}$ which can easily be transformed into Eq.~(\ref{monopole-density}).}:
\beq
\label{monopole-density}
\rho = 2^{-7} (1 \pm 0.02) \, e^6_3\,.
\eeq
At high temperatures we can use the dimensional reduction formalism and express 3D coupling constant
$e_3$ in terms of 4D Yang--Mills coupling $g$. At the tree level one has
\beq
\label{e3}
e^2_3 (T) = g^2(\Lambda,T) \, T\,,
\eeq
where $g(\Lambda,T)$ is the running coupling calculated at the scale
$T$,
\beq
\label{e4}
g^{-2}(\Lambda,T) ~=~ {11\over 12\pi^2} \log \Bigl({T \over \Lambda} \Bigr)~+~
{17\over44 \pi^2} \log \Bigr[2 \log \Bigl({T \over \Lambda}\Bigr)\Bigr]\,,
\eeq
and $\Lambda$ is a dimensional constant which can be determined  from lattice simulations.

At present, the lattice measurements of the $\Lambda$ parameter are ambiguous
and depend on the quantity which is used to determine it. We use two "extreme" values of $\Lambda$.
Namely in Ref.~\cite{Heller} the lattice data for the gluon propagator have been used to determine the
so--called "magnetic mass" in high temperature $SU(2)$ gluodynamics.
These measurements indirectly provide,
\beq
\label{lh}
\Lambda = 0.262(18)\, T_c \,,
\eeq
where $T_c$ is the temperature of the deconfinement phase transition,
$T_c \approx 0.69 \sqrt{\sigma}$ (see, e.g., \cite{Teper}). In
Ref.~\cite{Bali}, on the other hand, a the spatial string tension has been calculated and
a smaller value of $\Lambda$ was found:
\beq
\label{lb}
\Lambda = 0.076(13) \, T_c\,.
\eeq
Another uncertainty in our prediction is that we have used the dimensional
reduction which is supposed to work well only at
asymptotically high temperatures. In practice, however, the
temperatures used in lattice measurements are only few times larger
$T_c$.

%====================================================================

\subsection{Monopole potential in Abelian projection}

So far we assumed that the heavy monopole potential is measured via
the 't Hooft loop in full, non-Abelian lattice simulations.
However, within the Abelian dominance model all
the interactions are described by QED-like interactions of the
magnetic and electric charges. Moreover, the magnetic currents in the vacuum
can be measured directly. Then the potential energy of external sources
can, in principle, be evaluated using the ensembles of the magnetic currents.
In case of external color charges, or Wilson loop such
an approach is well known.
Also, the intermonopole potential has been studied on the lattice both at
zero~\cite{suganuma} and non--zero~\cite{stack} temperatures.
In this section we address this issue anew and emphasize that
in fact calculation of the heavy monopole potential involves
extra parameters. However, the situation is simplified greatly in
the limits of very low and very high temperatures.

The Abelian monopole action in the Maximal Abelian projection depends on many parameters. The action
contains the term $S^{(Coul)}[\dual j] = {\mathrm{const.}} (\dual
j,\Delta^{-1} \dual j)$, responsible for the Coulomb exchange between
monopoles as well as various $n$--point, $n\ge 2$, local monopole interaction
terms, Ref.~\cite{suzuki}. Here $\Delta^{-1}$ is the lattice inverse
Laplacian, $\dual j$ is the monopole current on the dual lattice.
Throughout this Section we are using the differential form notations on
the lattice~\cite{diff}.

Let us consider the following simple monopole action with 2--point (self) interaction as
an example:
\beqn
S^{(mon)}[\dual j] = 4 \pi^2 \beta (\dual j,\Delta^{-1} \dual j) + 4 \pi^2 \gamma 
{||\dual j||}^2\,,
\label{Sj}
\eeqn
This action corresponds to the London limit of the 
(dual) lattice Abelian Higgs
model~\cite{smit} in which the role of the Higgs field is played by the monopole field.
Another representation of model \eq{Sj} is the compact gauge field
representation~\cite{suzuki}:
\beqn
\cZ & = & \int\limits^{\pi}_{-\pi}
\dD \theta e^{- S^{(comp)}(\dd \theta)}\,, \nonumber\\
e^{- S^{(comp)}(\dd \theta)} & = &
\sum\limits_{n(c_2) \in \Z} \exp\Bigl\{ - \beta {||\dd \theta + 2 \pi n||}^2 - \gamma(\dd
\theta + 2 \pi n, \Delta (\dd \theta + 2 \pi n)) \Bigr\}\,,
\eeqn
where $\theta$ is the compact gauge field and $n$ is the integer valued auxiliary plaquette
field. The intermonopole potential can be studied then with the help of the Abelian 't~Hooft loop,
\beqn
H^{\mathrm{ab}}_{\Sigma_J,q} = \exp\{S^{(comp)}(\dd \theta) - S^{(comp)}(\dd \theta
+ 2 \pi q \, \dual \Sigma_J)\}\,,
\eeqn
where the surface $\Sigma_J$ ends on the trajectory of an external monopole with charge $q$.

Similarly to eq.~\eq{energy} the quantum average of the `t~Hooft operator gives
the static monopole potential:
\beqn
V^{\mathrm{ab}}_{m \bar m,q} (R) = - \frac{1}{T} \ln \frac{\cZ^{mon}_q}{\cZ^{mon}_0}\,,
\quad
\cZ^{mon}_q = \sum\limits_{\stackrel{\dual j(\dual c_1)\in \Z }{ \delta \dual j = 0}}
e^{ - S^{(mon)}[\dual j + q \, \dual J]}\,,
\eeqn
where the summation is taken over all closed monopole trajectories. The 2--point interaction term
provides the local interaction between external $J$ and dynamical $j$ currents, $S_{int} (\dual j,
\dual J) = 8 q \pi^2 \gamma (\dual j, \dual J)$. In our case the external current consists of two
disconnected pieces, $\dual J = \dual J_m + \dual J_{\dual m}$, separated by the distance $R$.

\begin{center}
\begin{figure}[!htb]
 \begin{minipage}{160mm}
\begin{tabular}{cc}
\epsfig{file=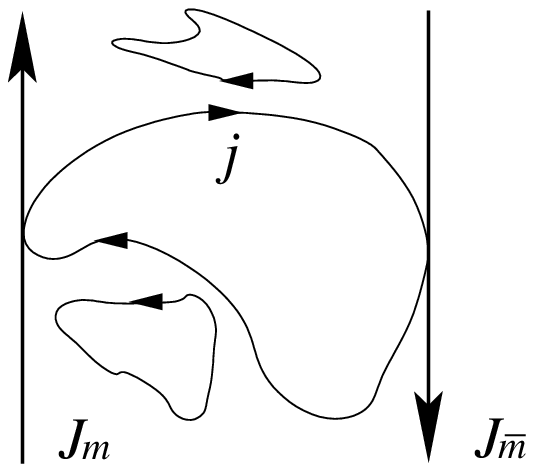,width=6.8cm} \hspace{1cm}&
\hspace{1cm}\epsfig{file=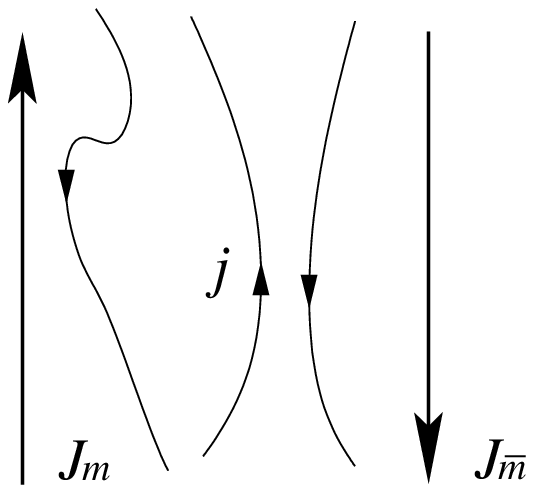,width=6.8cm}\\
(a) & (b) \\
 \end{tabular}
  \end{minipage}
\caption{The static external currents $J_m$ and $J_{\bar m}$ and the dynamical 
monopole currents $j$ at confinement, (a), and the deconfinement, (b), phases.}
\label{fig:j}
\end{figure}
\end{center}

The term $S_{int}(j,J)$ is non-zero iff the dynamical monopole current touches the external monopole current
$J$, see Figure~\ref{fig:j}(a). Thus, the $2$--point interaction terms affect the intermonopole potential
provided the dynamical monopoles overlap locally with the external sources.

Let us discuss the role of the local $n$--point interaction terms at the finite
temperature. At sufficiently small temperatures the system is in the confinement phase in
which the entropy of the monopole currents prevails over their energy. Thus there exist
arbitrarily long dynamical monopole currents $j$ connecting the external (anti-) monopole
trajectories $J_m$ ($J_{\bar m}$), Figure~\ref{fig:j}(a). In other words, the monopole
currents in the confinement phase of gluodynamics are percolating~\cite{ivanenko}: the
probability for two distinct points to be connected by a monopole trajectory does not
depend on the distances between them at sufficiently large separations. Since the dynamical
currents are not getting suppressed as their length is increased the distance dependent
part of the inter-monopole potential is not affected by the $n$--point interaction terms.

As temperature increases the system goes into the deconfinement phase. Now the $n$--point
terms become important since the monopole currents acquire non--zero free energy per unit
monopole trajectory length. However at sufficiently high temperatures the dynamical monopole
trajectories become static ($i.e.$, wrapped on the compactified direction with a minimal
length), see Fig.~\ref{fig:j}(b), and $n$--point terms do not affect the distance dependence of the
correlators for well separated external monopoles. 

To summarize we expect that the magnetic screening mass obtained using
only the long range part of the action should be correct at
asymptotically small and large temperatures while at the intermediate
temperatures the result is modified by the local current interactions.

%====================================================================
\subsection{Comparison with the data}

Data on $\vm$ are only beginning to accumulate and we summarize briefly
the comparison of the lattice measurements with predictions above:

{\it Short distances.} The Coulomb-like potential (\ref{small-R-potential})
is confirmed in the numerical simulations in the classical approximation
\cite{Hoelbling:2000su,gubarev1}.
There is no running of $g^2$ on this level of course. As for the full quantum
simulations the normalization (\ref{small-R-potential}) of the potential at short distances
is confirmed within a factor of about 2, Ref.~\cite{forcrand-1}.

As for the power corrections, all the data so far \cite{Hoelbling:2000su,forcrand} are
fitted smoothly with a Yukawa potential (\ref{yukawa}). Thus, there is no evidence
whatsoever that the linear term at short distances is to be subtracted. However, no explicit
bound on the linear term at short distances has been obtained either.

{\it Screening mass at $T=0$.} Existing data \cite{Hoelbling:2000su,forcrand} seem to be not
accurate
enough to distinguish between the models (\ref{glueball}), (\ref{predictionhiggs}).
Moreover, no checks of the vector-exchange have been made.

\begin{figure*}[!htb]
\centerline{\epsfig{file=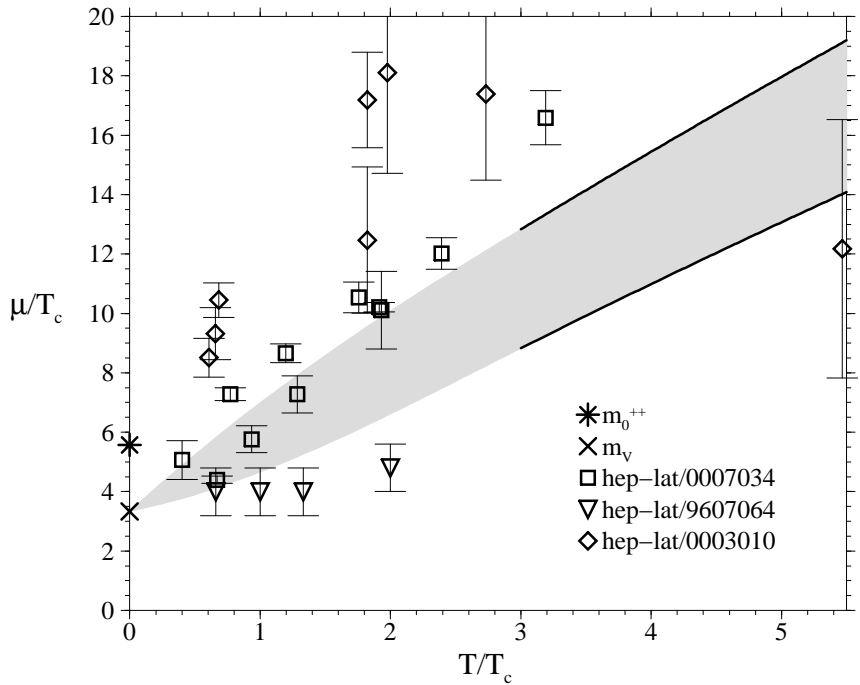,width=0.8\textwidth}}
\caption{Temperature dependence of the screening mass in the heavy
monopole potential (\ref{yukawa}).  Shaded region denotes the
theoretical prediction, Eqs.~(\ref{predictionhiggs}),
(\ref{m_D}-\ref{e4}).  Symbols denote:  ${\tt \times}$ -- the dual
gluon mass $m_V \approx 1 GeV$
as reported in Ref.~\cite{ilgenfritz}; {\tt *} -- the mass
of the lightest $0^{++}$ glueball \cite{Teper};
${\bigtriangledown}$ -- magnetic screening mass measured in
Maximal Abelian gauge, Ref.~\cite{stack}; $\Diamond$ and $\Box$ --
the screening masses measured at various temperatures in
Ref.~\cite{Hoelbling:2000su} and Ref.~\cite{forcrand} respectively.
}
\label{masses}
\end{figure*}

{\it Temperature dependence of the screening mass.} On Figure~\ref{masses} we have summarized the
existing data on temperature dependence of the screening mass together with our predictions,
Eqs.~(\ref{predictionhiggs}-\ref{e4}). The direct measurement of the screening mass
\cite{Hoelbling:2000su,forcrand} is shown by the diamonds and squares, respectively. The
character of the predicted temperature dependence is reproduced by the data. Although
numerically the screening mass seems to be systematically higher than is predicted, the
prediction based on the Abelian dominance hypothesis is rather close to the numerical
results
within the errors.

The triangles on Figure~1 denote the results of Ref.~\cite{stack} obtained in the Maximal
Abelian projection using the method described in the previous Section. At small temperatures
these results are quantitatively in agreement with Refs.~\cite{Hoelbling:2000su,forcrand}
while at higher temperatures the screening mass obtained in
Ref.~\cite{stack} falls essentially lower than is indicated
by the measurements in the full gluodynamics \cite{Hoelbling:2000su,forcrand}.
It is worth emphasizing that the calculations \cite{stack} took into account only the
Coulomb-like interaction between the monopoles. As is explained in the preceding
section, this is justified at very low and very high temperatures, but not at the
intermediate temperatures. This observation allows to understand why the results
of \cite{Hoelbling:2000su,forcrand} and \cite{stack} agree at $T=0$ and tend to disagree at
intermediate temperatures.

\section*{Conclusion}

To summarize, the heavy monopole potential at small distances is of the Coulomb type with a known overall
normalization. The large--distance potential is of the Yukawa type. The descent of the interaction at large
distances and at non--zero temperatures is qualitatively consistent with the monopole dominance model. Thus
the monopole dominance model of SU(2) gluodynamics based on the dual superconductivity predicts the inter
monopole potential which is consistent with the numerical data both at small and large distances at high as
well as at low temperatures. Further data are desired to allow for a detailed quantitative comparison.

%============================================
\section*{Acknowledgments}

Authors are thankful to Ph.~de~Forcrand for providing us with lattice data.
M.N.Ch. and M.I.P. acknowledge the kind hospitality of the staff of the
Max-Planck Institut f\"ur Physik (M\"unchen), where the work was initiated.
Work of M.N.C. and M.I.P. was partially supported by RFBR
99-01230a and Monbushu grants, and CRDF award RP1-2103.

%============================================

\end{document}